\documentclass[journal]{IEEEtran}
\usepackage[T1]{fontenc}
\usepackage[latin9]{inputenc}
\usepackage{amsmath}
\usepackage{amssymb}
\usepackage{amsthm,bm}
\usepackage{url}
\usepackage{cite}
\usepackage{babel,verbatim,balance}
\usepackage[rgb]{xcolor}
\usepackage{graphics, graphicx}
\usepackage{acronym}
\usepackage{upgreek}
\usepackage[bookmarks,colorlinks]{hyperref}
\usepackage{setspace,gensymb}

\acrodef{cir}[CIR]{Channel Impulse Response}

\begin{document}

\title{Reconfigurable Intelligent Surfaces for Rich Scattering Wireless Communications: Recent Experiments, Challenges, and Opportunities}
\author{
George C. Alexandropoulos,~\IEEEmembership{Senior~Member,~IEEE}, Nir Shlezinger,~\IEEEmembership{Member,~IEEE}, and Philipp del Hougne
\thanks{G.~C.~Alexandropoulos is with the Department of Informatics and Telecommunications, National and Kapodistrian University of Athens, 15784 Athens, Greece (e-mail: alexandg@di.uoa.gr).
}
\thanks{
N.~Shlezinger is with the School of Electrical and Computer Engineering, Ben-Gurion University of the Negev, Be'er-Sheva, Israel (e-mail: nirshl@bgu.ac.il).
}
\thanks{
P.~del~Hougne is with University of Rennes, CNRS, IETR - UMR 6164, F-35000, Rennes, France (e-mail: philipp.delhougne@gmail.com).}
}

\maketitle
\begin{abstract}
Recent advances in the fabrication and experimentation of Reconfigurable Intelligent Surfaces (RISs) have motivated the concept of the smart radio environment, according to which the propagation of information-bearing waveforms in the wireless medium is amenable to programmability. Although the vast majority of recent experimental research on RIS-empowered wireless communications gravitates around narrowband beamforming in quasi-free space, RISs are foreseen to revolutionize wideband wireless connectivity in dense urban as well as indoor scenarios, which are usually characterized as strongly reverberant environments exhibiting severe multipath conditions. In this article, capitalizing on recent physics-driven experimental explorations of RIS-empowered wave propagation control in complex scattering cavities, we identify the potential of the spatiotemporal control offered by RISs to boost wireless communications in rich scattering channels via two case studies. First, an RIS is deployed to shape the multipath channel impulse response, which is shown to enable higher achievable communication rates. Second, the RIS-tunable propagation environment is leveraged as an analog multiplexer to localize non-cooperative objects using wave fingerprints, even when they are outside the line of sight. Future research challenges and opportunities in the algorithmic design and experimentation of smart rich scattering wireless environments enabled by RISs for sixth Generation (6G) wireless communications are discussed.
\end{abstract}

\section{Introduction}
The recently proposed concept of the smart radio environment envisions fully programmable Electro-Magnetic (EM) propagation of information-bearing waveforms, which can be jointly optimized with transceiver signal processing~\cite{di2019smart_arxiv}. A key technological enabler for this vision is the usage of Reconfigurable Intelligent Surfaces (RISs) \cite{huang2019holographic_arxiv,hardware2020icassp,DMA2020}, which are artificial planar structures with nearly passive integrated electronic circuits whose reflection characteristics can be programmed to manipulate an incoming EM wave in a wide variety of functionalities \cite{huang2019reconfigurable_arxiv}. 
Precursors of the underlying concepts began to emerge in the early $2000$s~\cite{sievenpiper2003two,holloway2005reflection}, followed by another wave of early works around $10$ years later~\cite{subrt2012intelligent,Kaina_metasurfaces_2014,CQW14}.
The potential of RISs to boost wireless communications in terms of coverage extension, energy efficiency, and localization/positioning capability has lately been foreseen to play an important role in future sixth Generation (6G) wireless communications
\cite{Samsung,WavePropTCCN}.

Recent experimental research on RIS-empowered wireless communications focuses on the ability of RISs to generate controllable beam steering profiles in scenarios with poor scattering~\cite{tang2020wireless_arxiv,dai2020reconfigurable_arxiv}. By using design and analysis tools based on quasi-free-space assumptions, it has been also demonstrated that RISs can assist in overcoming harsh non Line-Of-Sight (LOS) conditions. Nonetheless, the smart radio environment concept envisions the deployment of RISs in indoor and dense urban setups, where the quasi-free-space assumption may not hold, and the channel is faithfully characterized by rich scattering models. In principle, the amount of scattering depends both on the radio propagation environment (e.g., its material properties and geometry) and the operating frequency band. In general, the higher the frequency is, the less scattering occurs. However, for operating frequencies below $6$~GHz, as for example WiFi and typical cellular frequencies, many wireless communication scenarios take place in highly reverberant environments. Such environments appear in dense urban network deployments as well as inside vessels, planes, train/metro stations, malls, and even in rooms inside buildings.

The application of RISs in rich scattering environments requires a different approach than in quasi-free space, due to the fundamentally different nature of EM wave propagation which is characterized by strong multipath. In quasi-free space, wireless communications rely on directive beamforming. The transmitted information-bearing waveforms are captured by the receiver at a feasibly estimated angle-of-arrival and time intervals. However, in rich scattering environments, each scatterer acts like a random secondary source and the receiver captures a multitude of copies of the emitted waveform at seemingly random time instants from seemingly random angles of incidence. Often, there may not even exist a direct LOS path between the transmitter and receiver. These differences have two important consequences. First, due to the long \ac{cir}, the transceiver scheme needs to account for the presence of Inter-Symbol Interference (ISI), which impacts signal processing complexity and achievable rate performance. Second, the receiver cannot localize the transmitter based on the angle or time of arrival of the transmitted waveform. In fact, receivers often perform pilot-assisted estimation of the channel coefficients at every channel coherence time, which requires designated effort that is deprived from the desired data communication. The latter consequences, together with the fundamental differences between the RIS application in quasi-free-space and rich scattering setups, motivate the theoretical and experimental investigation of the potential, capabilities, and algorithmic design methods for RISs in empowering wireless communications in rich scattering environments.

In this article, we capitalize on recent physics-driven experimental explorations of EM wave control in RIS-coated complex scattering cavities \cite{dupre2015wave,del2016spatiotemporal_arxiv,del2018precise_arxiv} to experimentally identify the potential and key challenges of RIS-empowered wireless communication in rich scattering environments. Specifically, we present two experimental case studies. The first case study deploys an RIS prototype to shape the long multipath \ac{cir} at wish, demonstrating its ability to yield improved rate performance. The second case study utilizes the RIS-tunable propagation environment as an analog multiplexer to localize non-cooperative objects on a grid via Wave FingerPrinting (WFP). We conclude with a discussion of open research challenges and opportunities for RIS-empowered communications realized in rich scattering scenarios.

\section{RISs for Rich Scattering Wireless Channels} 
In this section, we present RIS design guidelines for rich scattering wireless communications, and discuss their fundamental differences from scenarios of free-space EM signal propagation. We also present an \textit{in situ} characterization method for RIS prototypes in rich scattering environments and apply it to an RIS design used in subsequent experiments.

\subsection{Design Guidelines}\label{subsec:Guidleines}
An RIS, also termed programmable metasurface, is a planar array of ultra-thin meta-atoms (also known as cells or elements), each with multiple digitalized states corresponding to distinct EM responses. For quasi-free-space beam manipulation, akin to the main scenario currently considered for RISs in the wireless communications literature~\cite{tang2020wireless_arxiv,dai2020reconfigurable_arxiv}, a fine-grained control over the reflected EM field is essential for accurate beamforming. This fact motivated researchers to rely on meta-atoms of \textit{sub-wavelength} size (e.g., $\lambda/10$ with $\lambda$ being the wavelength~\cite{huang2019holographic_arxiv}), despite inevitable strong mutual coupling \cite{ESPARs2014globecom} among meta-atoms, and well-defined gray-scale-tunable EM properties.

In contrast, in rich scattering environments, the wave energy is statistically equally spread throughout the wireless medium and the ensuing \textit{ray chaos} implies that rays impact the RIS from all possible, rather than one well-defined, directions. 
Instead of creating a directive beam, the goal becomes the manipulation of as many ray paths as possible. This manipulation may either aim at tailoring those rays to create constructive interference at a target location (Case Study~I in Section~\ref{sec:Equalization}) or to efficiently stir the field (Case Study II in Section~\ref{sec:Localization}).   
These manipulations can be efficiently realized with \textit{half-wavelength-sized} meta-atoms, which enable the control of more rays with a fixed amount of electronic components (PIN diodes, etc.), as compared to RISs equipped with their sub-wavelength counterparts. Additionally, mutual coupling among half-wavelength meta-atoms is weaker, if not negligible.   
Interestingly, given the chaotic nature of the rays, the specific EM response of a meta-atom is irrelevant, as long as different states significantly impact the rays. Furthermore, the additional benefits of gray-scale over one-bit tunability are often outweighted by the increased complexity of the electronic circuitry.

\begin{figure}
\centering
\includegraphics [width =  \columnwidth]{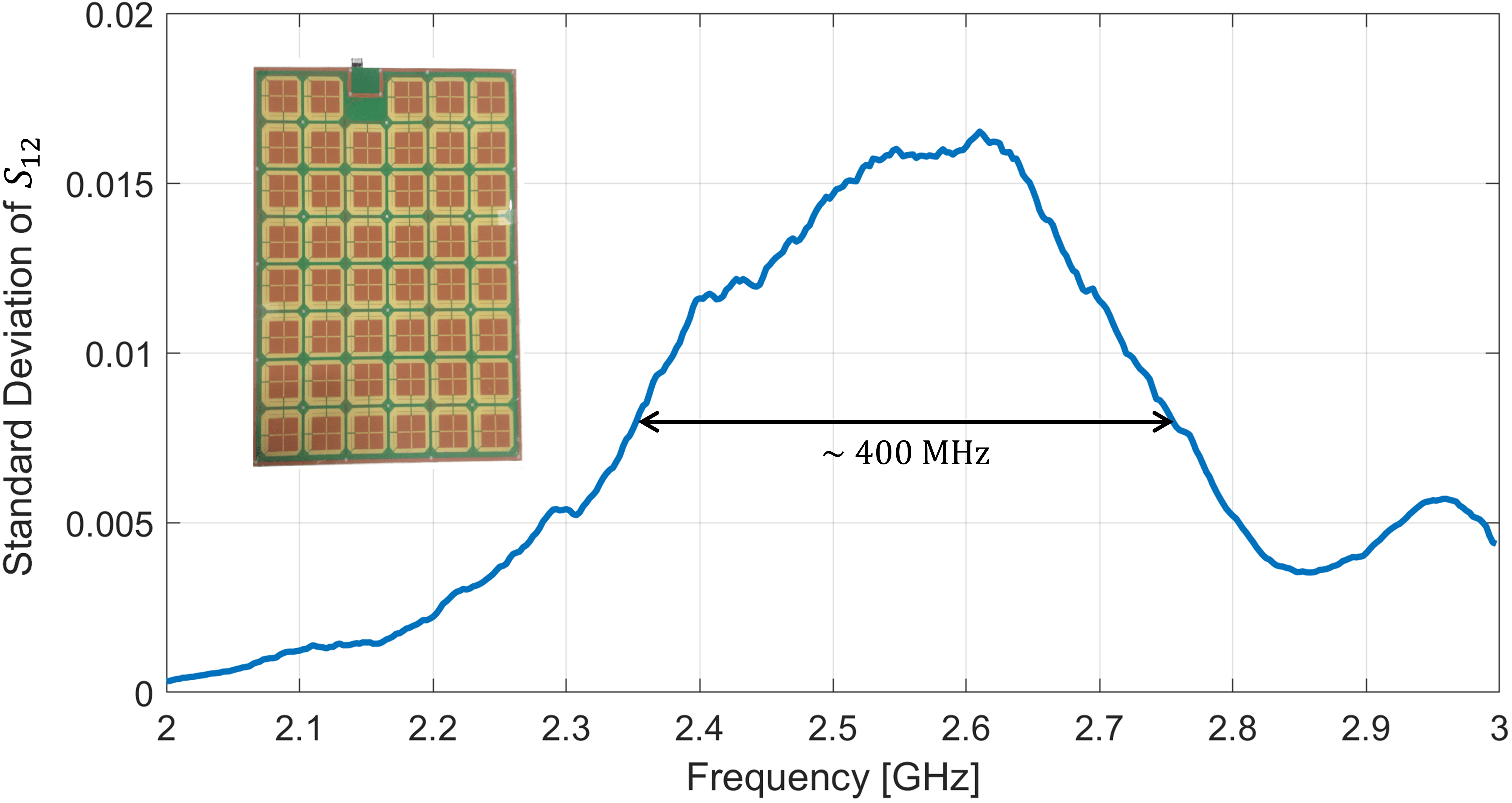}
\caption{\textit{In situ} characterization of a $47$-element RIS prototype consisting of one-bit programmable meta-atoms (see inset), as characterized in the experimental setup of Fig.~\ref{fig4}. The curve quantifies the ability of the RIS to manipulate the $S_{12}$ parameter between two monopole antennas in the given setting, accounting for all possible angles of incidence and polarizations.}
\label{fig1}
\end{figure}


\subsection{\textit{In Situ} Characterization of an Experimental Prototype}
\label{subsec:Prototype}
The EM responses of RISs are often characterized in terms of the reflection coefficients of an isolated meta-atom, under normal incidence in its various states. This approach neglects the complexity of the EM problem, in which the angle of incidence and polarization of the impinging wave as well as the mutual coupling among adjacent meta-atoms and the effects of nearby scatterers all impact the EM responses. Such simplifications can be warranted if the RIS acts as reflectarray, as part of the transmit architecture~\cite{dai2020reconfigurable_arxiv}. In a rich scattering  environment, however, waves from all possible angles and with all possible polarizations are incident on the RIS. Characterizing the meta-atom response as a function of the incidence angle, polarization, nearby scatterers, and other factors is impractical, and even unnecessary, because ray tracing is unfeasible in any case. Indeed, the subsequent two case studies will showcase that RISs in rich scattering environments can be either optimized iteratively by trial and error or used with random meta-atom configurations. All that matters is the extent to which reconfiguring the meta-atoms impacts in some way the scattering properties of the propagation environment. A practical technique to assess the ability of an RIS to manipulate the EM field inside a given rich scattering environment is to measure \textit{in situ} the $S_{12}$ parameter between one transmitting and one receiving antenna for a series of random RIS configurations. The standard deviation across these measurements constitutes a useful metric to assess the extent to which the RIS is capable of impacting the transmission under consideration. Besides the RIS's surface area and its position relative to the transceivers, the impact of the RIS on the EM field also depends on the amount of reverberation: the longer the waves propagate inside the medium, the more often they interact with the RIS and are affected by its configuration.

We illustrate the aforedescribed \textit{in situ} characterization technique in Fig.~\ref{fig1} for an RIS prototype consisting of one-bit and half-wavelength-sized programmable meta-atoms, deployed directly in the targeted multipath setup of our second case study (see Fig.~\ref{fig4}). We emphasize that all results discussed in this paper are expected to qualitatively hold irrespective of the specific employed RIS design. The utilized prototypes follow the design proposed in~\cite{kaina2014hybridized_arxiv}. Each meta-atom consists of two resonators that hybridize: by controlling the electrical length of one resonator via the bias voltage of a PIN diode, the overall EM response can be chosen to be on or off resonance at the working frequency around 2.5~GHz, resulting in two distinct states for which the desired phase of the reflection coefficient differs by roughly $\pi$ under normal incidence~\cite{kaina2014hybridized_arxiv}. The meta-atoms illustrated in the photographic image in Fig.~\ref{fig1} can be understood as the fusion of two meta-atoms from~\cite{kaina2014hybridized_arxiv}, one rotated by $90^\circ$, each acting on one EM field polarization. We plot in Fig.~\ref{fig1} the standard deviation over 500 random RIS configurations of the transmission measured \textit{in situ} between the two monopole antennas, as seen in Fig.~\ref{fig4}. This metric quantifies the impact of the RIS on the transmission at each considered frequency. It is apparent that the RIS most efficiently manipulates the EM waves in the vicinity of 2.5~GHz, within a $400$~MHz bandwidth. This bandwidth limitation originates from the fact that the meta-atom design is based on a resonant effect.

\section{Case Study I: Analog Multipath Shaping}\label{sec:Equalization}
\label{subsec:EqualizationBackground}
One of the main characteristics of wireless communications in rich scattering environments is multipath propagation. The CIR resulting from multipath components has a non-negligible temporal support, thus inducing ISI, which is typically handled by dedicated transceiver signal processing methods. Receivers operating in multipath channels commonly utilize pre-processing and equalization filters, as means of mitigating the lengthy CIRs, potentially affecting the shape and level of the thermal noise \cite[Ch. 11]{goldsmith2005wireless}. On the transmitter side, transmission schemes such as Orthogonal Frequency Division Multiplexing (OFDM), which are known to be suitable for multipath channels, must implicitly account in their design for the \ac{cir} in terms of its coherence bandwidth. This characteristic dictates the subcarrier spacing, as well as its temporal support, via the length of the cyclic prefix separating successive OFDM blocks. 

The adoption of RISs in multipath channels brings forth the possibility to externally modify their scattering profile. A complex scattering enclosure mixes spatial and temporal degrees of freedom, and thus, the purely spatial control of the RIS enables shaping the channel in both \textit{space} and \textit{time}. Specifically, by controlling the phase of some rays, constructive interference at the receiver at a desired moment in time can be created. Spatial focusing results in increasing the effective Signal-to-Noise Ratio (SNR) at the specific physical location of the receiver, akin to traditional beam steering. On the other hand, temporal focusing implies that the RIS can result in the receiver observing qualitatively different \acp{cir} compared to those imposed by the original physical wireless channel, and possibly applying a form of \textit{analog equalization}, without shaping the thermal noise present at the receiver side. 

In the following, we present an experimental study showcasing the ability of an RIS to externally shape the multipath profile of rich scattering wireless environments. We first present the experimental setup, followed by a numerical evaluation of how the achieved analog channel shaping translates into an achievable rate performance improvement.

\subsection{Experimental Setup and RIS Optimization}
\begin{figure}[!t]
\centering
\includegraphics [width =  \columnwidth]{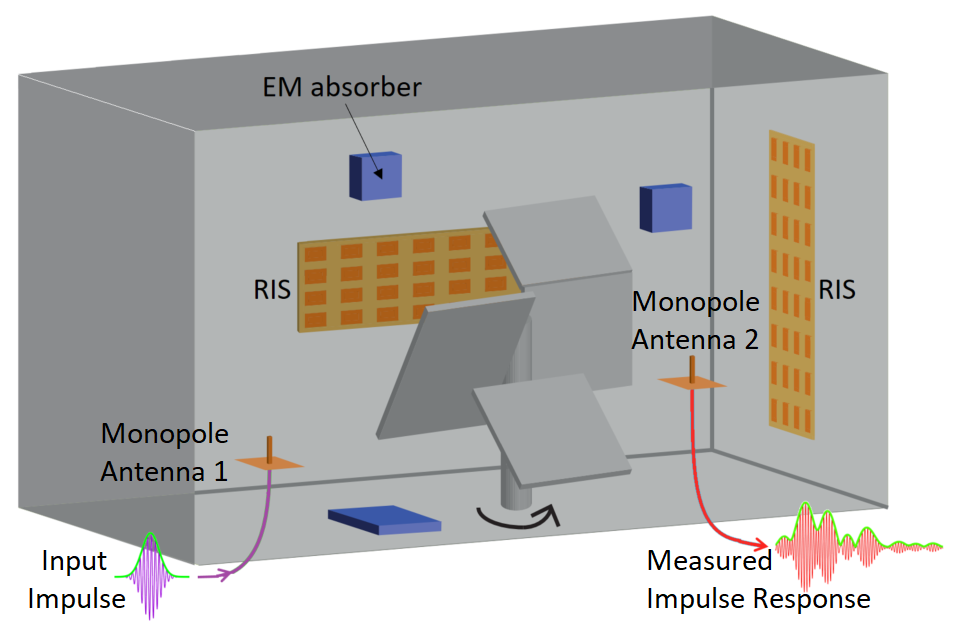}
\caption{Experimental setup for Case Study I (analog multipath shaping based on~\cite{del2016spatiotemporal_arxiv}). The walls of a reverberant irregular metallic enclosure are partially covered by a $102$-element spatially distributed RIS with one-bit programmable meta-atoms. The RIS enables shaping the highly complex SIR between two monopole antennas.}
\label{fig2}
\end{figure}

\begin{figure*}
	\centering
\includegraphics[width=0.8\linewidth]{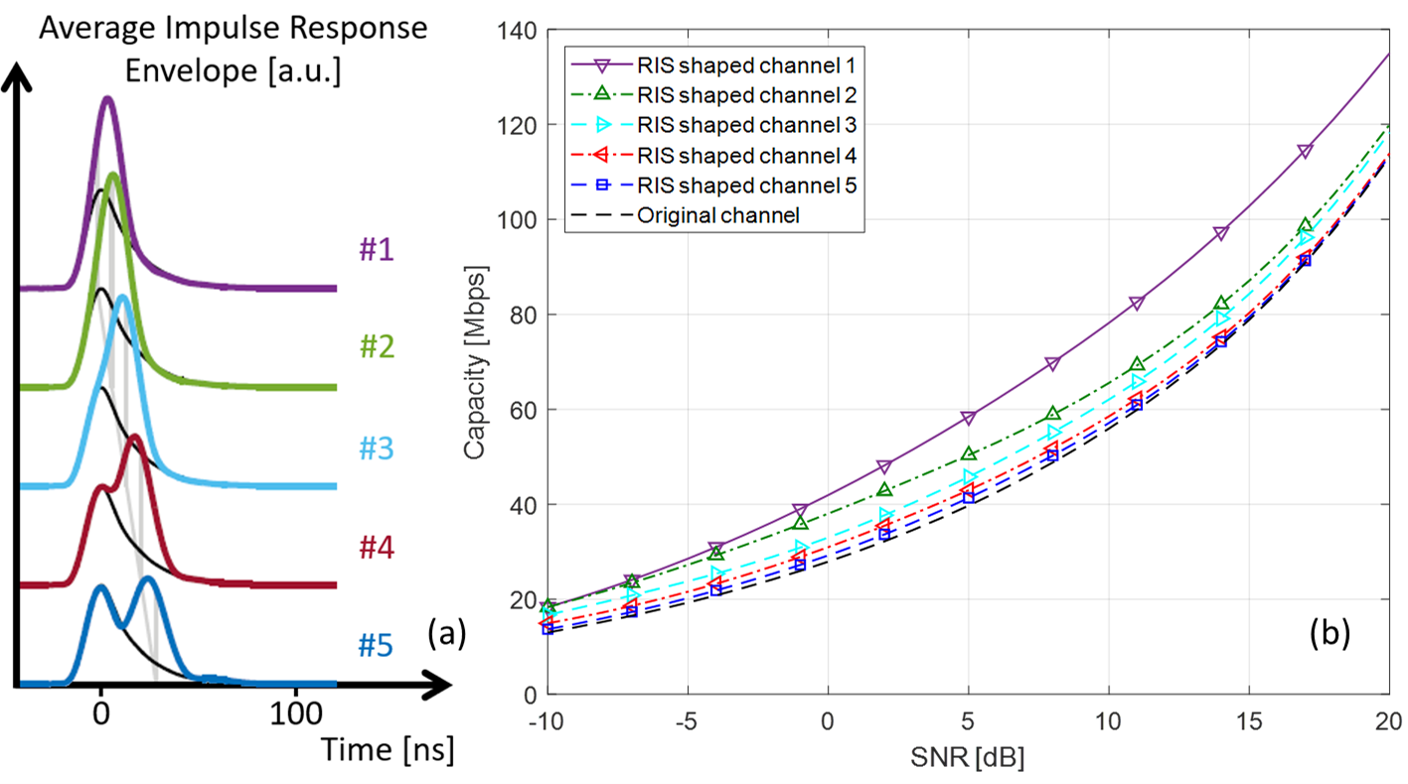}
\caption{Experimental results for RIS-based analog multipath shaping: $(a)$ measured \acp{cir} averaged over $60$ realizations based on~\cite{del2016spatiotemporal_arxiv}, where each curve (thick colored lines) compares the \ac{cir} shaping for a different temporal focusing setting (focal time indicated by a thin gray line) to the original CIR (thin black line); $(b)$ capacity of the RIS-shaped channels in the presence of additive white Gaussian noise as a function of the SNR in dB.}
	\label{fig3}
\end{figure*}

The experimental setup sketched in Fig.~\ref{fig2} consists of an irregular metallic enclosure of volume 1.1~$\text{m}^3$ and total surface 6.6~$\text{m}^2$~\cite{del2016spatiotemporal_arxiv}. Around the operating frequency at 2.5~GHz, this environment is very reverberant: upon excitation with a 66~MHz pulse, the average \ac{cir} duration is more than three times longer than the emitted pulse, as shown in the black curves in Fig.~\ref{fig3}(a). A spatially distributed RIS with $102$ elements in total, where each meta-atom offers one-bit control over the reflected phase along one polarization, partially covers two of the enclosure's walls.

In the experiment, the RIS configuration was optimized to maximize the \ac{cir} envelope at a desired temporal position along the \ac{cir}, as indicated in gray in Fig.~\ref{fig3}(a). The measured \ac{cir} was averaged over $60$ realizations of the disorder by rotating a large irregular metallic structure inside the enclosure. Given the disordered nature of the environment, no analytical or numerical tools to identify the optimal RIS configuration are available. The impact of the RIS on the channel is highly complex, and is very difficult to accurately model analytically. Instead, the RIS configuration was optimized iteratively: starting from a random configuration, the state of one meta-atom was flipped, the new channel was measured, and the change was kept if the desired channel characteristic (the envelope magnitude at the desired temporal position) was improved. The procedure was looped multiple times over each meta-atom, because the optimal configuration of a given meta-atom is to some extent correlated with the configuration of the other meta-atoms due to the reverberation.

The measured example results of the RIS-shaped \acp{cir} are depicted in Fig.~\ref{fig3}(a). These results demonstrate that by properly configuring the RIS prototype, one can effectively shape the multipath scattering profile, and thus the transfer function of the wireless channel. For instance, by assigning different RIS configurations, one can obtain \acp{cir} varying from an impulse-like channel, as the shaped channel $1$ in Fig.~\ref{fig3}(a), to a bi-modal impulse response, e.g., the shaped channel $5$ in the same subfigure. 


\subsection{Capacity Comparisons of RIS-Shaped Channels}

In contrast to any form of signal processing operation performed solely at the receiver (e.g., equalization) or transmitter that cannot alter the capacity of a given wireless propagation channel, the RIS modifies the \ac{cir}, implying its potential for increasing the channel capacity. Moreover, unlike filtering carried out upon reception, shaping the \ac{cir} using the RIS does not affect the thermal noise at the receiver. To illustrate the potential for capacity improvements, we evaluate in the sequel the achievable data rates for the different RIS-shaped \ac{cir}s seen in Fig.~\ref{fig3}(a).

We numerically evaluate the capacity of ISI channels with additive white Gaussian noise, in which the CIRs are given by the averaged envelopes in Fig.~\ref{fig3}(a). In particular, in Fig.~\ref{fig3}(b), we compare the numerically evaluated capacity of the five averaged RIS-shaped channels seen in Fig.~\ref{fig3}(a) (colored curves), which was computed via waterfilling in the frequency domain \cite[Ch. 4.3]{goldsmith2005wireless}, to the corresponding capacity of the averaged channel envelope without RIS; the latter is the black curve in Fig.~\ref{fig3}(a). We evaluate the capacities over a $66$~MHz band of the RIS-equalized channels versus the SNR in dB, which was defined as the inverse of the noise variance. Since the capacity measures for the ISI channels are computed for the averaged CIRs of Fig.~\ref{fig3}(a), the curves in Fig.~\ref{fig3}(b) constitute an upper bound on the ergodic capacity over those random multipath fading channels \cite[Ch. 4]{goldsmith2005wireless}. It is shown in this figure that, by tuning the RIS to shape the \ac{cir}, the data rates that one can reliably achieve over such channels can be notably improved. For instance, the time-coherent focusing of the energy of the \ac{cir}, as carried out by the first two RIS-based equalized channels, results in significant capacity improvements. 
These results indicate that the capability of a properly configured RIS to externally shape the \ac{cir} of multipath channels can significantly improve the achievable throughput and reliability of wireless communications in rich scattering environments. 

\section{Case Study II: Localization via \\EM Wave Fingerpints}

\label{sec:Localization}
Precise localization using Radio-Frequency (RF) signals is of paramount importance for fifth Generation (5G) and future 6G wireless communications~\cite{patwari2005locating,6G_Flagship_arxiv,leitinger2019belief,mendrzik2019enabling,li2019massive,wymeersch2019radio,nearfieldRIS2021icc}. Personal navigation, human-machine interaction, immersive technologies, indoor positioning, ambient-assisted living, security monitoring, radar sensing, asset tracking, and various other vertical applications of wireless sensor networks and internet-of-things require precise localization. In general, when accurate external positioning data is unavailable, accurate location estimations facilitate network management, initial access, and transceiver position tracking. For example, in millimeter wave communications, the availability of the receiver's position estimation can improve the rate and latency performance of initial access and beam tracking. Localization has been traditionally based on ray tracing, with the simplest example being the triangulation method. State-of-the-art methods rely on information such as time-of-arrival or angle-of-arrival to localize objects~\cite{witrisal2016high_arxiv}.
However, localization via ray-tracing techniques becomes intractable in rich scattering environments, such as the one depicted in Fig.~\ref{fig4}, due to the severe multipath effects. Conventional ray tracing cannot discriminate between the objects of interest and other strong reflectors. So far, the leading approach to combat this complexity is to encircle the region of interest with a dedicated wireless sensor network to perform statistical shadowing analysis. 

In wave physics, it is well known that chaotic wave fields are extremely sensitive to tiny perturbations. This fact points toward a completely different approach to localization that \textit{makes the complexity a virtue}: the EM wave field can be seen as a fingerprint of the object position.  Irrespective of the presence of strong scatterers and the absence of a LOS component, each object position corresponds to a unique chaotic wave field. Conventional wisdom would suggest to rely on spatial or spectral diversity by employing an array of sensors or broadband measurements to obtain a vector of transmission measurements at different positions or frequencies, respectively, that is unique for each object position, and hence, can serve as WFP. Ideally, however, one could devise a single-frequency and single-port scheme to minimize the hardware and computational costs of the localization system as well as the spectrum allotment constraints. RISs make this possible by offering \textit{configurational diversity}: one can probe the EM wave field at a single position and a single frequency with a fixed series of random RIS configurations, yielding a measurement vector that can serve as WFP~\cite{del2018precise_arxiv}. Remarkably, this approach utilizes a random series of RIS configurations, which implies the absence of any need to optimize the RIS configuration in contrast to the Case Study I.

\subsection{Experimental Setup}
\begin{figure}[!t]
\centering
\includegraphics [width =  \columnwidth]{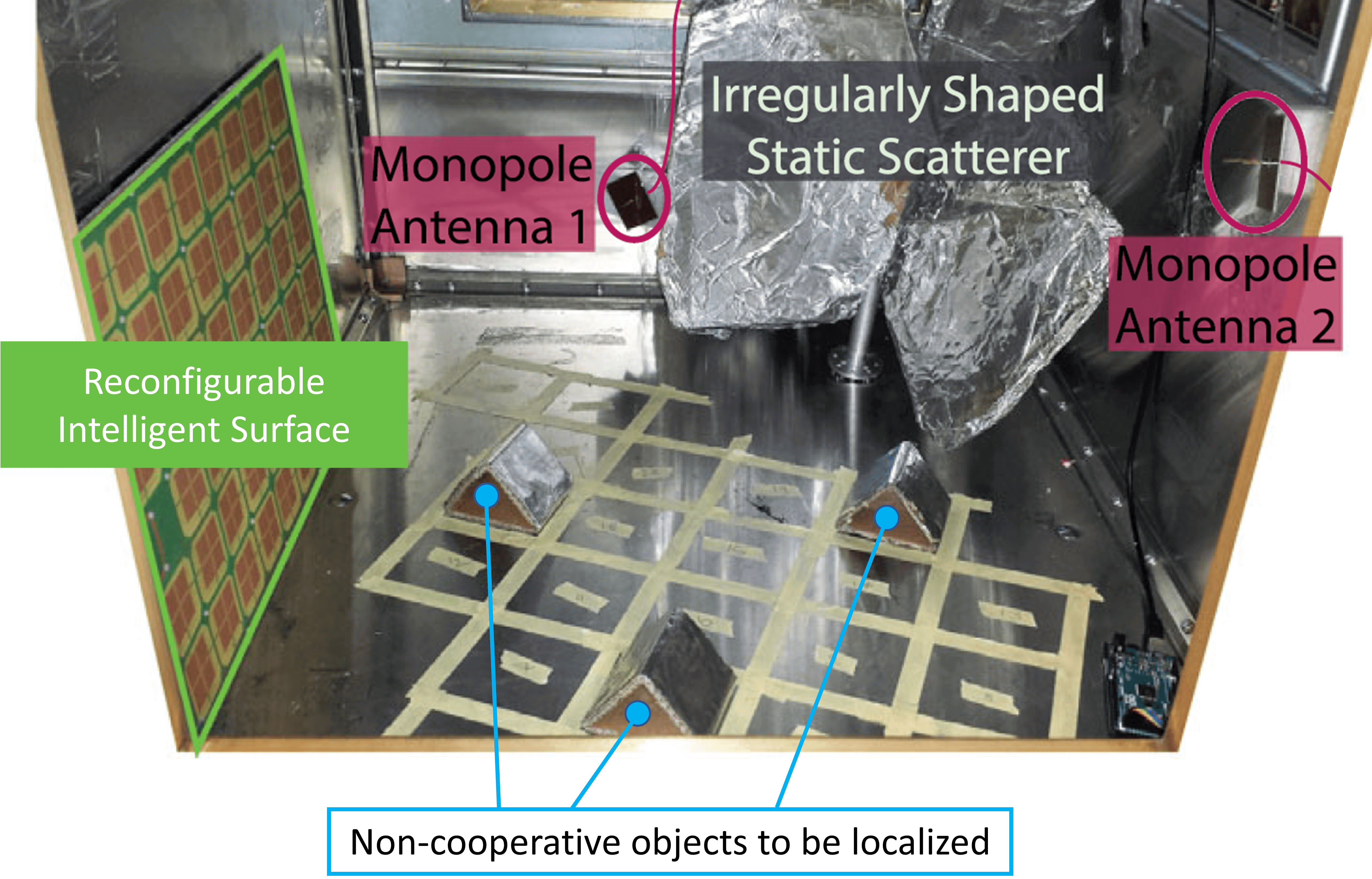}
\caption{Experimental setup for Case Study II (localization via EM wave fingerprints based on~\cite{del2018precise_arxiv}): three identical non-cooperative objects are to be localized inside a rich scattering and irregularly shaped metallic enclosure via the transmission between two monopole antennas and the RIS prototype characterized in Fig.~\ref{fig1}. The grid identifies the $23$ predefined object positions.}
\label{fig4}
\end{figure}
In the experimental demonstration inside an irregularly shaped metallic scattering enclosure of volume 1.1~$\text{m}^3$, as depicted in Fig.~\ref{fig4}, three identical triangular objects lying outside the LOS of two monopole antennas are to be localized~\cite{del2018precise_arxiv}. To showcase the potential of RISs in facilitating localization in such rich scattering environments, non-cooperative objects are considered, which neither emit beacon signals nor carry tags, although our WFP-based technique can also be applied to the localization of cooperative objects. This case refers to objects that are active network nodes, but do not implement any localization mechanism, thus saving power consumption and computational resources for other communication operations. The $47$-element RIS prototype depicted and characterized in Fig.~\ref{fig1} partially covers one of the enclosure walls, as shown in Fig.~\ref{fig4}. The WFPs are inked with transmission measurements between the two monopole antennas at a single frequency for a predefined, yet random, series of RIS configurations. 

Each of the three identical objects is allowed to occupy one of the $23$ predefined positions within the grid indicated on the enclosure's floor. We seek to identify the indices of the occupied positions; hence, the accuracy is evaluated in terms of a success rate between 0 and 1 (each correctly identified position index is a success), rather than a continuous position variable.
Given the rich scattering nature of the environment, it is impossible to selectively illuminate each grid position with a directive beam, and hence, solve the problem with 23 measurements. Instead, each possible combination of the three objects on the allowed position will yield a distinct measurement vector. However, it is unpractical to measure for calibration purposes all these ${23 \choose 3} = 1771$ measurement vectors, even in our small-scale demonstration. Instead, in the considered calibration step, only the WFPs corresponding to a single object were measured (i.e., $23$ instead of $1771$ calibration measurements) and it was assumed that they can be combined linearly for multiple objects. This assumption holds to a good approximation in the considered setup because inter-object scattering is almost negligible.

\subsection{Position Sensing Performance}
Given the measurement vector for an unknown distribution of objects, the processing challenge is to identify the combination of WFPs that best explains the measurements. One possibility is to do so by comparing the measurement against all possible combinations of WFPs~\cite{del2018precise_arxiv}. However, this approach is computationally very inefficient and scales unfavorably for problems with more objects and/or allowed positions. To overcome this issue, we have trained a simple fully-connected Artificial Neural Network (ANN) to predict the object positions, which used the measurement vector as input. The designed ANN has a single hidden layer of $256$ neurons with intermediate Rectified Linear Unit (ReLU) activation and $23$ output neurons followed by sigmoid activation. It is noted that the model-agnostic nature of ANNs allows them to learn such complex mappings without imposing a model on the underlying statistics that relate the measurements and the location of the objects. Considering that the complex scattering process in the experimental setup of Fig.~\ref{fig4} is likely to encode relevant scene features in long-range spatial structures, a fully-connected architecture appears better suited to the task, as compared to alternative ANN structures such as  convolutional neural networks, which are good at picking up local features.

\begin{figure} 
\centering
\includegraphics [width =  \columnwidth]{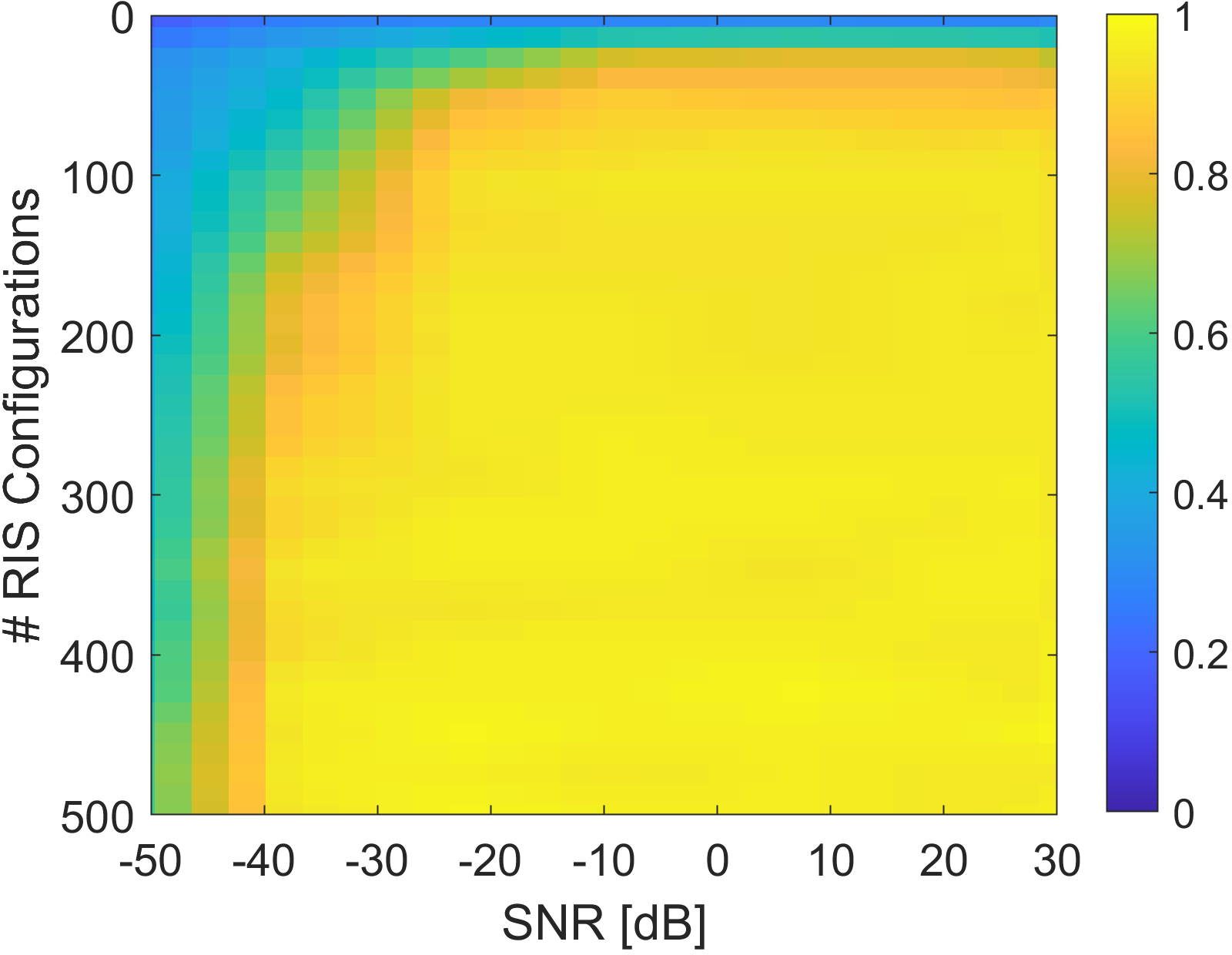} 
\caption{Experimental results for the proposed RIS-empowered WFP localization scheme demonstrating the fraction of the correctly identified object positions, as a function of the number of utilized random RIS configurations and the operating SNR in dB.}
\label{fig5}
\end{figure}

In Fig.~\ref{fig5}, we illustrate the localization accuracy of the proposed RIS-empowered scheme, using the experimental data measured in the setup of Fig.~\ref{fig4}, as a function of the SNR in dB and the length of the series of random RIS configurations utilized to ink the WFPs. The experimental SNR was 30~dB, and to obtain the figure we added synthetic white noise to explore lower SNR values. The results in Fig.~\ref{fig5} reveal the possibility of achieving an excellent localization accuracy, even at a very poor SNR (as low as $-40$~dB) or using as few as $100$ RIS configurations. We attribute these properties to the ANN's efficiency in extracting relevant information, even from very limited and/or noisy data. Being able to successfully apply the localization scheme at extremely low SNR values means that one can operate with very low levels of radiated power, improving energy efficiency and EM Field exposure metrics. The fact that as few as $100$ measurements were enough to distinguish between $1771$ configurations evidences the compressive nature of EM wave propagation in the RIS-controlled rich scattering environment.  

\section{Research Challenges and Opportunities}
\label{sec:Challenges}
A common idea emerging from our two case studies is the use of smart RIS-empowered propagation environments as \textit{analog processors}. In other words, computational tasks for wireless communications can be performed by RIS-controlled EM wave propagation, as opposed to using traditional electronic processors at the communication ends. In our first case study, the RIS prototype was used to externally adapt the \ac{cir} of the multipath channel, allowing to generate various forms of different \ac{cir} profiles. The ability to modify the \ac{cir}, without affecting the thermal noise at the receiver, allows RISs to increase the channel capacity. In the second case study, it was experimentally shown that an RIS facilitates RF-based localization in multipath settings. This was enabled by an ANN-aided compressive sensing scheme that multiplexes spatial object information across independent measurement modes generated by the RIS.

We envision that the analog processing role of RISs in smart rich scattering propagation environments can be taken even further in the future. For instance, the case study detailed in Section~\ref{sec:Equalization} focuses on the ability of RISs to shape a \ac{cir} impacting a single transmitter-receiver pair, where the RIS configuration is optimized to achieve a \ac{cir} with a desired temporal peak. The fact that such a setting was shown to improve capacity indicates that by designing RISs to shape multipath \acp{cir}, in light of a capacity maximization objective, it is expected to yield even more notable improvements. Furthermore, the abilities of RISs to jointly shape multiple \acp{cir} in multi-user communications under rich scattering environments, investigated to date only for single-frequency channels \cite{del2019optimally_arxiv}, and the resulting trade-offs between different users are left for future exploration. Last but not least, algorithmic designs that jointly optimize the RIS-enabled analog processing with the transceiver functionalities, requiring affordable computational complexity and control signaling overhead, are necessary. We also believe that analog multipath shaping will also impact wireless communication on the chip scale~\cite{imani2020toward}.

Additional opportunities follow from the usage of RISs to enhance the localization accuracy in rich scattering environments, which is \textit{not} fundamentally limited by the utilized wavelength~\cite{del2021deeply}.
As detailed in Section~\ref{sec:Localization}, the task of localizing objects or network nodes can be conveniently attributed to RISs, saving power consumption and computational resources for the transceivers. In addition, this functionality can be either jointly optimized with conventional localization schemes for scattering environments to further improve estimation accuracy, or used in conjunction with other communication functionalities that profit from location information (e.g., millimeter wave beam steering assisted by low-frequency localization). 
Moreover, by interpreting the RIS-programmable measurement process as a first trainable physical layer and the subsequent data analysis as additional trainable digital layers, it should be possible to optimize the utilized RIS configuration sequence, enabling the same performance with fewer measurements that contain more relevant information~\cite{del2020learned_arxiv}. Essentially, the optimized RIS configuration will pre-select information that is relevant for the localization task, already as part of the measurement process, thereby avoiding the indiscriminate acquisition of both useful and useless information. 
Further reducing the number of measurements necessary for precise localization will improve a wide range of metrics including latency, power consumption, processing burden, and EM field exposure.

Finally, given the extreme complexity of rich scattering environments that became apparent in the two presented case studies, the future potential of ANNs to capture these effects is evident. ANNs can learn complex mappings from data without relying on analytically tractable models, making them suitable to handling the inherent complexities associated with utilizing RISs in rich scattering setups. ANNs can be trained to capture the impact of the RIS on the channel, allowing the experimental iterative optimization procedure, used in the first case study, to be performed numerically. For the localization technique, on the one hand, ANNs capable of selecting statistically relevant information from noisy data, for instance, due to dynamically evolving environments (e.g., mobility conditions inducing Doppler or time selectivity), are an enticing perspective~\cite{LocalizationDynamicEnvironment}. This may be successful even if the parasitic perturbation is stronger than that induced by the objects to be localized. On the other hand, more elaborate ANNs may be capable of updating themselves in-field, taking advantage of the increasing amount of data that is accumulated during communication operation. It also appears feasible to transfer knowledge of a localization problem in one setting to a similar problem in another setting.

\section{Conclusion}
We explored the role of RISs in enhancing wireless communications under rich scattering environments, where EM wave propagation drastically differs from the usual assumptions of beam-like propagation. Based on recent experimental results from physics-driven explorations, we clarified design priorities for RISs in such scenarios and evaluated their application potential in two key case studies. First, it was shown that an RIS can shape at wish the lengthy CIRs found under severe multipath conditions. Second, the ability of an RIS to multiplex information about object locations onto a single detector was demonstrated to facilitate RF localization in strongly reverberant environments. We discussed the prominent role of RIS-enabled smart radio propagation environments as analog processors in both considered case studies and identified several future directions for this concept. We pointed out different ways in which artificial intelligence may play a dominant role in 6G wireless communications under rich scattering channels.

\bibliographystyle{IEEEtran}
\bibliography{references}

\begin{thebibliography}{10}
\providecommand{\url}[1]{#1}
\csname url@samestyle\endcsname
\providecommand{\newblock}{\relax}
\providecommand{\bibinfo}[2]{#2}
\providecommand{\BIBentrySTDinterwordspacing}{\spaceskip=0pt\relax}
\providecommand{\BIBentryALTinterwordstretchfactor}{4}
\providecommand{\BIBentryALTinterwordspacing}{\spaceskip=\fontdimen2\font plus
\BIBentryALTinterwordstretchfactor\fontdimen3\font minus
  \fontdimen4\font\relax}
\providecommand{\BIBforeignlanguage}[2]{{%
\expandafter\ifx\csname l@#1\endcsname\relax
\typeout{** WARNING: IEEEtran.bst: No hyphenation pattern has been}%
\typeout{** loaded for the language `#1'. Using the pattern for}%
\typeout{** the default language instead.}%
\else
\language=\csname l@#1\endcsname
\fi
#2}}
\providecommand{\BIBdecl}{\relax}
\BIBdecl

\bibitem{di2019smart_arxiv}
M.~Di~Renzo, M.~Debbah, D.-T. Phan-Huy, A.~Zappone, M.-S. Alouini, C.~Yuen,
  V.~Sciancalepore, G.~C. Alexandropoulos, J.~Hoydis, H.~Gacanin, J.~de~Rosny,
  A.~Bounceu, G.~Lerosey, and M.~Fink, ``Smart radio environments empowered by
  reconfigurable {AI} meta-surfaces: an idea whose time has come,''
  \emph{EURASIP J. Wireless Commun. Net.}, vol. 2019, no.~1, pp. 1--20, May
  2019.

\bibitem{huang2019holographic_arxiv}
C.~Huang, S.~Hu, G.~C. Alexandropoulos, A.~Zappone, C.~Yuen, R.~Zhang,
  M.~Di~Renzo, and M.~Debbah, ``Holographic {MIMO} surfaces for 6{G} wireless
  networks: Opportunities, challenges, and trends,'' \emph{IEEE Wireless
  Commun.}, vol.~27, no.~5, pp. 118--125, Oct. 2020.

\bibitem{hardware2020icassp}
G.~C. Alexandropoulos and E.~Vlachos, ``A hardware architecture for
  reconfigurable intelligent surfaces with minimal active elements for explicit
  channel estimation,'' in \emph{Proc. IEEE ICASSP}, Barcelona, Spain, May
  2020, pp. 9175--9179.

\bibitem{DMA2020}
N.~Shlezinger, G.~C. Alexandropoulos, M.~F. Imani, Y.~C. Eldar, and D.~R.
  Smith, ``Dynamic metasurface antennas for {6G} extreme massive {MIMO}
  communications,'' \emph{IEEE Wireless Commun. Mag.}, to appear, 2021.

\bibitem{huang2019reconfigurable_arxiv}
C.~Huang, A.~Zappone, G.~C. Alexandropoulos, M.~Debbah, and C.~Yuen,
  ``Reconfigurable intelligent surfaces for energy efficiency in wireless
  communication,'' \emph{IEEE Trans. Wireless Commun.}, vol.~18, no.~8, pp.
  4157--4170, Aug. 2019.

\bibitem{sievenpiper2003two}
D.~F. Sievenpiper, J.~H. Schaffner, H.~J. Song, R.~Y. Loo, and G.~Tangonan,
  ``Two-dimensional beam steering using an electrically tunable impedance
  surface,'' \emph{IEEE Trans. Antennas Propag.}, vol.~51, no.~10, pp.
  2713--2722, Oct. 2003.

\bibitem{holloway2005reflection}
C.~L. Holloway, M.~A. Mohamed, E.~F. Kuester, and A.~Dienstfrey, ``Reflection
  and transmission properties of a metafilm: With an application to a
  controllable surface composed of resonant particles,'' \emph{IEEE Trans.
  Electromagn. Compat.}, vol.~47, no.~4, pp. 853--865, Nov. 2005.

\bibitem{subrt2012intelligent}
L.~Subrt and P.~Pechac, ``Intelligent walls as autonomous parts of smart indoor
  environments,'' \emph{IET Commun.}, vol.~6, no.~8, pp. 1004--1010, May 2012.

\bibitem{Kaina_metasurfaces_2014}
N.~Kaina, M.~Dupr\'{e}, G.~Lerosey, and M.~Fink, ``Shaping complex microwave
  fields in reverberating media with binary tunable metasurfaces,'' \emph{Sci.
  Rep. 4}, pp. 1--7, Article No 076401, 2014.

\bibitem{CQW14}
T.~J. Cui, M.~Q. Qi, X.~Wan, J.~Zhao, and Q.~Cheng, ``Coding metamaterials,
  digital metamaterials and programmable metamaterials,'' \emph{Light Science
  Appl.}, vol.~3, no.~10, pp. e218--e218, 2014.

\bibitem{Samsung}
``The next hyper- {C}onnected experience for all,'' White Paper, Samsung 6G
  Vision, Jun. 2020.

\bibitem{WavePropTCCN}
G.~C. Alexandropoulos, G.~Lerosey, M.~Debbah, and M.~Fink, ``Reconfigurable
  intelligent surfaces and metamaterials: {T}he potential of wave propagation
  control for {6G} wireless communications,'' \emph{IEEE ComSoc TCCN
  Newslett.}, vol.~6, no.~1, pp. 25--37, Jun. 2020.

\bibitem{tang2020wireless_arxiv}
W.~Tang, M.~Z. Chen, X.~Chen, J.~Y. Dai, Y.~Han, M.~Di~Renzo, Y.~Zeng, S.~Jin,
  Q.~Cheng, and T.~J. Cui, ``Wireless communications with reconfigurable
  intelligent surface: Path loss modeling and experimental measurement,''
  \emph{IEEE Trans. Wireless Commun.}, vol.~20, no.~1, pp. 421--439, Jan. 2021.

\bibitem{dai2020reconfigurable_arxiv}
L.~Dai, B.~Wang, M.~Wang, X.~Yang, J.~Tan, S.~Bi, S.~Xu, F.~Yang, Z.~Chen,
  M.~Di~Renzo, and L.~Hanzo, ``Reconfigurable intelligent surface-based
  wireless communications: Antenna design, prototyping, and experimental
  results,'' \emph{IEEE Access}, vol.~8, pp. 45\,913--45\,923, Mar. 2020.

\bibitem{dupre2015wave}
M.~Dupr{\'e}, P.~del Hougne, M.~Fink, F.~Lemoult, and G.~Lerosey, ``Wave-field
  shaping in cavities: Waves trapped in a box with controllable boundaries,''
  \emph{Phys. Rev. Lett.}, vol. 115, no.~1, p. 017701, 2015.

\bibitem{del2016spatiotemporal_arxiv}
P.~del Hougne, F.~Lemoult, M.~Fink, and G.~Lerosey, ``Spatiotemporal wave front
  shaping in a microwave cavity,'' \emph{Phys. Rev. Lett.}, vol. 117, no.~13,
  p. 134302, 2016.

\bibitem{del2018precise_arxiv}
P.~del Hougne, M.~F. Imani, M.~Fink, D.~R. Smith, and G.~Lerosey, ``Precise
  localization of multiple noncooperative objects in a disordered cavity by
  wave front shaping,'' \emph{Phys. Rev. Lett.}, vol. 121, no.~6, p. 063901,
  2018.

\bibitem{ESPARs2014globecom}
G.~C. Alexandropoulos, V.~I. Barousis, and C.~B. Papadias, ``Precoding for
  multiuser {MIMO} systems with single-fed parasitic antenna arrays,'' in
  \emph{Proc. IEEE GLOBECOM}, Austin, USA, Dec. 2014, pp. 3656--3661.

\bibitem{kaina2014hybridized_arxiv}
N.~Kaina, M.~Dupr{\'e}, M.~Fink, and G.~Lerosey, ``Hybridized resonances to
  design tunable binary phase metasurface unit cells,'' \emph{Opt. Express},
  vol.~22, no.~16, pp. 18\,881--18\,888, 2014.

\bibitem{goldsmith2005wireless}
A.~Goldsmith, \emph{Wireless Communications}.\hskip 1em plus 0.5em minus
  0.4em\relax Cambridge University Press, 2005.

\bibitem{patwari2005locating}
N.~Patwari, J.~N. Ash, S.~Kyperountas, A.~O. Hero, R.~L. Moses, and N.~S.
  Correal, ``Locating the nodes: cooperative localization in wireless sensor
  networks,'' \emph{IEEE Signal Process. Mag.}, vol.~22, no.~4, pp. 54--69,
  Jul. 2005.

\bibitem{6G_Flagship_arxiv}
A.~Bourdoux, A.~N. Barreto, B.~van Liempd, C.~de~Lima, D.~Dardari, and
  D.~Belot, ``{6G} white paper on localization and sensing,'' \emph{[Online]
  arxiv.org/abs/2006.01779}, Jun. 2020.

\bibitem{leitinger2019belief}
E.~Leitinger, F.~Meyer, F.~Hlawatsch, K.~Witrisal, F.~Tufvesson, and M.~Z. Win,
  ``A belief propagation algorithm for multipath-based {SLAM},'' \emph{IEEE
  Trans. Wirel. Commun.}, vol.~18, no.~12, pp. 5613--5629, Dec. 2019.

\bibitem{mendrzik2019enabling}
R.~Mendrzik, F.~Meyer, G.~Bauch, and M.~Z. Win, ``Enabling situational
  awareness in millimeter wave massive {MIMO} systems,'' \emph{IEEE J. Sel.
  Topics Signal Process.}, vol.~13, no.~5, pp. 1196--1211, Sep. 2019.

\bibitem{li2019massive}
X.~Li, E.~Leitinger, M.~Oskarsson, K.~{\AA}str{\"o}m, and F.~Tufvesson,
  ``Massive {MIMO}-based localization and mapping exploiting phase information
  of multipath components,'' \emph{IEEE Trans. Wireless Commun.}, vol.~18,
  no.~9, pp. 4254--4267, Sep. 2019.

\bibitem{wymeersch2019radio}
H.~Wymeersch, J.~He, B.~Denis, A.~Clemente, and M.~Juntti, ``Radio localization
  and mapping with reconfigurable intelligent surfaces,'' \emph{IEEE Veh.
  Technol. Mag.}, vol.~15, no.~4, pp. 52--61, Dec. 2020.

\bibitem{nearfieldRIS2021icc}
Z.~Abu-Shaban, K.~Keykhosravi, M.~F. Keskin, G.~C. Alexandropoulos,
  G.~Seco-Granados, and H.~Wymeersch, ``Near-field localization with a
  reconfigurable intelligent surface acting as lens,'' in \emph{Proc. IEEE
  ICC}, Montreal, Canada, Jun. 2021.

\bibitem{witrisal2016high_arxiv}
K.~Witrisal, P.~Meissner, E.~Leitinger, Y.~Shen, C.~Gustafson, F.~Tufvesson,
  K.~Haneda, D.~Dardari, A.~F. Molisch, A.~Conti, and M.~Z. Win,
  ``High-accuracy localization for assisted living: {5G} systems will turn
  multipath channels from foe to friend,'' \emph{IEEE Signal Process. Mag.},
  vol.~33, no.~2, pp. 59--70, Mar. 2016.

\bibitem{del2019optimally_arxiv}
P.~del Hougne, M.~Fink, and G.~Lerosey, ``Optimally diverse communication
  channels in disordered environments with tuned randomness,'' \emph{Nat.
  Electron.}, vol.~2, no.~1, pp. 36--41, 2019.

\bibitem{imani2020toward}
M.~F. Imani, S.~Abadal, and P.~del Hougne, ``Toward dynamically adapting
  wireless intra-chip channels to traffic needs with a programmable
  metasurface,'' in \emph{in Proc. ACM NanoCoCoA}, 2020, pp. 20--25.

\bibitem{del2021deeply}
M.~del Hougne, S.~Gigan, and P.~del Hougne, ``Deeply sub-wavelength
  localization with reverberation-coded-aperture,'' \emph{[Online]
  arxiv.org/abs/2102.05642}, 2021.

\bibitem{del2020learned_arxiv}
P.~del Hougne, M.~F. Imani, A.~V. Diebold, R.~Horstmeyer, and D.~R. Smith,
  ``Learned integrated sensing pipeline: Reconfigurable metasurface
  transceivers as trainable physical layer in an artificial neural network,''
  \emph{Adv. Sci.}, vol.~7, no.~3, p. 1901913, 2020.

\bibitem{LocalizationDynamicEnvironment}
P.~del Hougne, ``Robust position sensing with wave fingerprints in dynamic
  complex propagation environments,'' \emph{Phys. Rev. Research}, vol.~2, p.
  043224, 2020.

\end{thebibliography}

\end{document}